\documentclass[fleqn,usenatbib]{mnras}

\usepackage{txfonts}

\usepackage[T1]{fontenc}

\DeclareRobustCommand{\VAN}[3]{#2}
\let\VANthebibliography\thebibliography
\def\thebibliography{\DeclareRobustCommand{\VAN}[3]{##3}\VANthebibliography}

\usepackage{graphicx}	
\usepackage{natbib}
\usepackage{url}
\usepackage{longtable}
\usepackage{threeparttable}
\usepackage{ulem}
\usepackage{footnote}
\usepackage[section]{placeins}

\newcommand{\source}{4U~1820$-$30}

\newcommand{\arcsecond}{$^{\prime\prime}$}

\newcommand{\uJy}{$\mu$Jy}

\newcommand{\uJybeam}{$\mu$Jy\,beam$^{-1}$}
\newcommand{\maxi}{\textsl{MAXI}}
\newcommand{\rxte}{\textsl{RXTE}}

\newcommand{\ergscm}{erg\,s$^{-1}$\,cm$^{-2}$}
\newcommand{\swift}{\textit{Swift}}

\newcommand{\ergs}{erg\,s$^{-1}$}

\newcommand{\nh}{$N_{\mathrm{H}}$}

\newcommand{\phscm}{photons\,cm$^{-2}$\,s$^{-1}$}

\graphicspath{{./}}

\title[The evolving radio jet in 4U~1820$-$30]{The evolving radio jet from the neutron star X-ray binary 4U~1820$-$30}

\author[T. D. Russell et al.]{
T. D. Russell,$^{1,2}$\thanks{E-mail:thomas.russell@inaf.it}
N. Degenaar,$^{2}$
J. van den Eijnden,$^{3}$
M. Del Santo,$^{1}$
A. Segreto,$^{1}$
D. Altamirano,$^{4}$
\newauthor 
A. Beri,$^{4}$
M. D\'{i}az Trigo,$^{5}$
J.~C.~A. Miller-Jones$^{6}$\\
$^{1}$INAF, Istituto di Astrofisica Spaziale e Fisica Cosmica, Via U. La Malfa 153, I-90146 Palermo, Italy\\
$^{2}$Anton Pannekoek Institute for Astronomy, University of Amsterdam, Science Park 904, NL-1098 XH Amsterdam, The Netherlands\\
$^{3}$Astrophysics, Department of Physics, University of Oxford, Denys Wilkinson Building, Keble Road, Oxford OX1 3RH, UK\\
$^{4}$School of Physics and Astronomy, University of Southampton, Highfield SO17 IBJ, England\\
$^{5}$ESO, Karl-Schwarzschild-Strasse 2, 85748 Garching bei München, Germany\\
$^{6}$International Centre for Radio Astronomy Research - Curtin University, GPO Box U1987, Perth, WA 6845, Australia
}

\date{Accepted 23-July-2021. Received 23-July-2021; in original form 23-July-2021}

\pubyear{2021}

\begin{document}
\label{firstpage}
\pagerange{\pageref{firstpage}--\pageref{lastpage}}
\maketitle

\begin{abstract}
The persistently bright ultra-compact neutron star low-mass X-ray binary \source\ displays a $\sim$170\,d accretion cycle, evolving between phases of high and low X-ray modes, where the 3 -- 10\,keV X-ray flux changes by a factor of up to $\approx 8$. The source is generally in a soft X-ray spectral state, but may transition to a harder state in the low X-ray mode. Here, we present new and archival radio observations of \source\ during its high and low X-ray modes. For radio observations taken within a low mode, we observed a flat radio spectrum consistent with \source\ launching a compact radio jet. However, during the high X-ray modes the compact jet was quenched and the radio spectrum was steep, consistent with optically-thin synchrotron emission. The jet emission appeared to transition at an X-ray luminosity of $L_{\rm X (3-10 keV)} \sim 3.5 \times 10^{37} (D/\rm{7.6\,kpc})^{2}$\,\ergs. We also find that the low-state radio spectrum appeared consistent regardless of X-ray hardness, implying a connection between jet quenching and mass accretion rate in \source, possibly related to the properties of the inner accretion disk or boundary layer.

\end{abstract}




\section{Introduction}

There is an observed phenomenological relationship between jet production and the accretion process in all accreting compact objects, however, at present the exact nature of the coupling remains to be determined. Neutron star (NS) and black hole (BH) low mass X-ray binary (LMXB) systems act as useful observational laboratories to study this coupling as they allow for the properties of the in-flowing accretion flow and out-flowing jets to be observed on human timescales. Changes in the accretion flow are typically observed in the X-rays, while corresponding changes in the jet emission are generally seen in the radio band. In particular, NS LMXBs offer an important observational tool allowing for the role that stellar surface, spin, and magnetic field play on jet production to be tested. However, due to the relative radio faintness of NS LMXBs \citep[e.g.,][]{2018MNRAS.478L.132G}, our understanding of jet launching in NS LMXBs is generally less evolved than their BH counterparts \citep[e.g.,][]{2006MNRAS.366...79M}.

Jets from accreting NS LMXBs show a diverse range of behaviour \citep[e.g.,][]{2006MNRAS.366...79M,2011IAUS..275..233M,2017MNRAS.470..324T, 2017MNRAS.470.1871G,2018MNRAS.478L.132G,2018ApJ...869L..16R,2020MNRAS.492.2858G}, and there has been some debate as to whether NS LMXB jets show a similar pattern of behaviour to BH systems \citep[see, e.g.,][]{2006MNRAS.366...79M,Marino20}. NS and BH LMXBs can launch two types of jets depending on the state of the source: a steady compact jet or a bright transient jet \citep[e.g.,][]{2006MNRAS.366...79M}, depending on the mass accretion rate ($\dot{m}$) and structure of the accretion flow. At lower $\dot{m}$ during the harder X-ray spectral states the compact jet is characterised by optically-thick synchrotron emission exhibiting a flat to inverted radio spectrum ($\alpha \sim 0$, where the radio flux density, $S_{\nu}$, is proportional to the observing frequency, $\nu$, such that $S_{\nu} \propto \nu^{\alpha}$; \citealt{1979ApJ...232...34B}). At higher $\dot{m}$, when the source is softer, the compact jet emission can be quenched \citep{2003MNRAS.342L..67M,2009MNRAS.400.2111T,2010ApJ...716L.109M,2017MNRAS.470.1871G} and a transient jet is launched, although jet quenching is sometimes not observed \citep{1998ATel....8....1R,2003A&A...399..663K,2004MNRAS.351..186M,2011IAUS..275..233M}. Recent results from BH LMXBs suggest that jet quenching at radio frequencies results as the jet spectral break, which is the break from optically-thick to optically-thin synchrotron emission for the highest energy synchrotron spectra, evolves below the radio band \citep[e.g.,][]{2020MNRAS.498.5772R}, implying that the jet acceleration region shifted away from the compact object \citep[e.g.,][]{2001A&A...372L..25M,2005ApJ...635.1203M}. A similar evolution of the jet spectral break is thought to occur in NS systems \citep{2018A&A...616A..23D}. Around this time the short-lived, rapidly-flaring transient jet can be launched, exhibiting a steep radio spectrum ($\alpha \sim -0.7$) originating from ejected discrete (optically-thin) synchrotron emitting plasma \citep[e.g.,][]{2006csxs.book..381F}. Later, as the accretion rate reduces the source transitions back to the hard state and the compact jet re-establishes.

\source{} is an ultra-compact NS LMXB located in the globular cluster NGC~6624, at $\approx 7.6$\,kpc \citep{2003A&A...399..663K}. This persistently X-ray bright object displays a $\sim$170\,d accretion cycle \citep[e.g.,][]{1984ApJ...284L..17P,2001ApJ...563..934C}, where the X-ray luminosity evolves between L$_{\rm low} \sim 8 \times 10 ^{36} (D/\rm{7.6\,kpc})^{2}$\,erg\,s$^{-1}$ and L$_{\rm high} \sim 6 \times 10 ^{37} (D/\rm{7.6\,kpc})^{2}$\,erg\,s$^{-1}$ (3--10\,keV) \citep[e.g.,][]{1984ApJ...284L..17P,2003A&A...405..199S,2006ApJS..163..372W}, which we refer to as low and high X-ray modes, respectively. Generally in a soft X-ray state, \source\ can sometimes transition to a hard state at its lowest mass accretion rates \citep[during low modes, e.g.,][]{2007ApJ...654..494T,2013ApJ...767..160T}.

Here, we present broadband radio observations of \source\ taken during its high and low X-ray modes. Observations and analysis are given in Section 2. Section 3 reports the results from the radio observations and connecting the observed changes to the accretion flow. Conclusions are in Section 4.

\section{Observations}
\label{sec:obs}

\subsection{Radio}

All radio results are given in Table~\ref{tab:radio}1.

\vspace{-4mm}

\subsubsection{VLA}
\label{sec:our_VLA}

In total, we present eight radio observations of \source\ from the Karl G. Jansky Very Large Array (VLA). The VLA observed \source\ on 2018-12-09 (MJD~58461; project code 18A-194) and 2020-07-07 (MJD~59037; project code 20A-255), with data taken at L-band (1--2\,GHz), C-band (4.5--5.5 and 7--8\,GHz), and K-band (18--19 and 25--26\,GHz), where all bands were observed within a 2-hour window but not simultaneously. For these radio observations 3C286 was used for bandpass and flux calibration, and J1820$-$2528 was used for phase calibration. The VLA also observed NGC~6624 five times between 2018 May 22 -- 28 (MJDs\,58260, 58261, 58262, 58265, and 58266) under project code 18A-081 \citep[][]{2020ApJ...903...73S}. Those data were recorded at X-band (8 -- 12\,GHz), with durations of 1 or 2\,hours. Flux and bandpass calibration were done using either 3C286 or 3C48 and phase calibration used J1820$-$2528. Due to flaring of 3C48 at the time of these VLA observations a 10\% uncertainty was added in quadrature to the measured errors for those epochs.

Using the Common Astronomy Software Application (\textsc{casa} version 5.1.2; \citealt{2007ASPC..376..127M}), data were edited for radio frequency interference (RFI) and systematics, before being calibrated following standard procedures. L-band and C-band observations were imaged with a Briggs robust parameter of 0 to balance sensitivity and resolution. K- and X-band data were imaged with natural weighting to maximise sensitivity. To characterise the radio spectral behaviour the L- and C-band observations were separated into $4 \times 256$ and $4 \times 512$\,MHz sub-bands, respectively. The X-band observations were separated into 4$\times$1\,GHz sub-bands. The radio flux density of the target was determined by fitting for a point source in the image plane using a Gaussian with a full width half maximum set to the synthesised beam of the observation.

\subsubsection{ATCA}

The Australia Telescope Compact Array (ATCA) observed NGC~6624 on 2006-07-05 (MJD 53921), 2006-07-23 (MJD 53939), and 2006-08-08 (MJD~53955) under project code C1559. These observations were recorded simultaneously at 4.80 and 8.64\,GHz, with 104\,MHz of bandwidth at each central frequency. ATCA also observed the target field on 2015-04-24 (MJD~57136) and 2015-04-25 (MJD~57137) under project code C2877 \citep[see][Tudor et al. in prep.]{2018ApJ...862...16T}. These two later observations were carried out at central frequencies of 5.5 and 9\,GHz, and recorded using the Compact Array Broad-band Backend (CABB; \citealt{2011MNRAS.416..832W}) providing an increased bandwidth of 2\,GHz at each central frequency. All ATCA observations used PKS~1934$-$638 for primary calibration and 1822$-$36 for phase calibration. Data were edited, calibrated, and imaged (using a natural weighting) as described above.

\begin{figure*}
\centering
\includegraphics[width=0.9\textwidth]{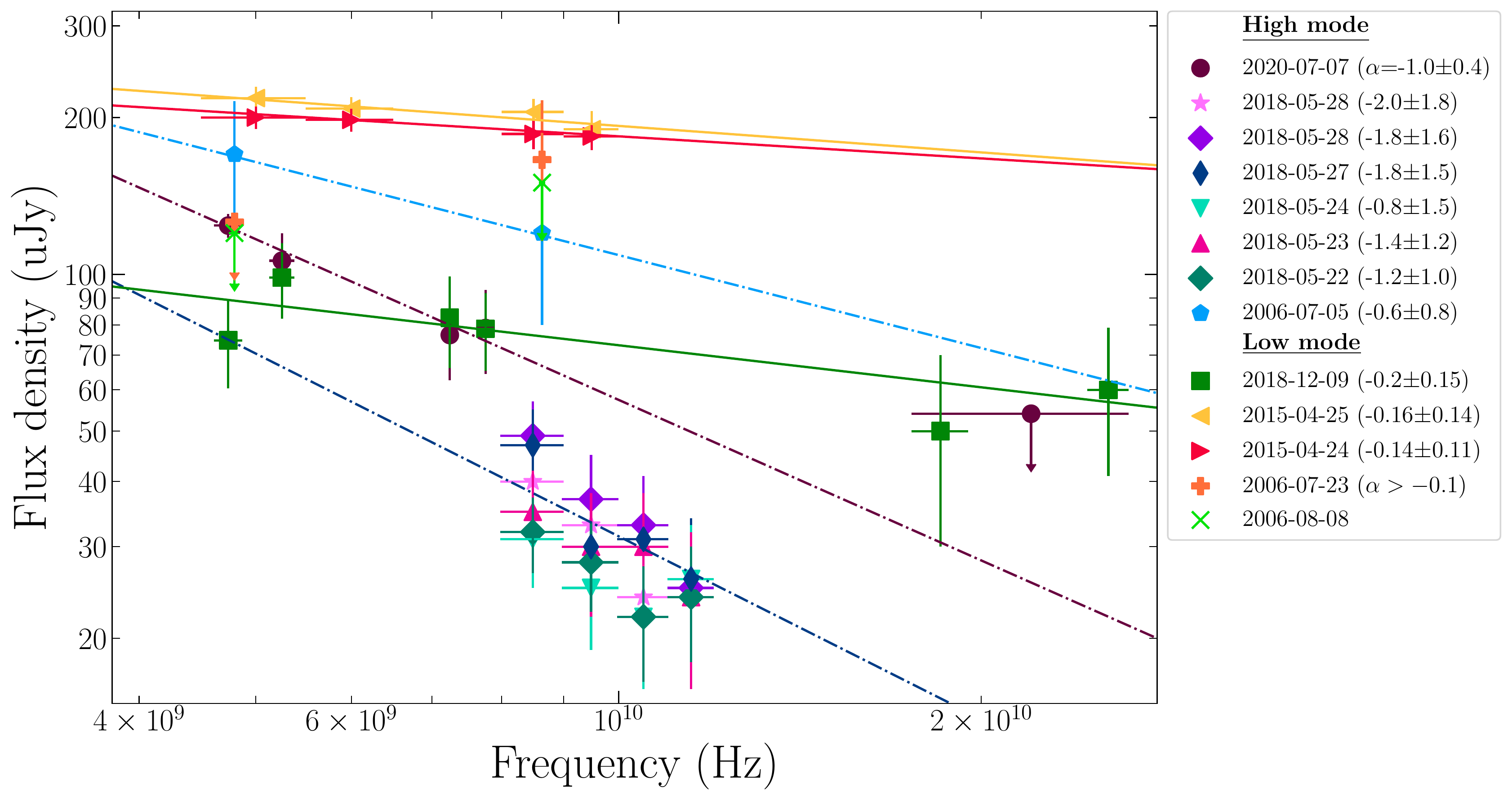}
\caption{Radio emission from \source\ during its high and low X-ray modes. All data are provided in Table~\ref{tab:radio}1. Fits to each epoch are shown, where dash-dotted lines were used for radio observations taken during high X-ray modes and solid lines were used for low X-ray modes. Radio spectral indices ($\alpha$) for each epoch are provided in the legend. Due to the radio results showing little difference for the 2018 high mode data, for clarity, we only show the fit to these observations stacked together (which gives $\alpha = -1.32 \pm 0.54$). Non-detections are shown as three-sigma upper limits (determined as three times the image noise). For all observations, when the source was in a high X-ray mode the radio spectrum appeared steep ($\alpha \approx -1$), while during low X-ray modes the radio spectrum was flat ($\alpha \approx -0.15$), indicating a dramatic change in the jet between the X-ray modes.}
\label{fig:radio_xrb}
\end{figure*}

\subsection{X-ray}
\subsubsection{\maxi, \swift-BAT, and \rxte-ASM}
Daily count rates in the 2--20\,keV band were obtained from the Monitor of All-sky X-ray Image (\maxi) X-ray telescope \citep{2009PASJ...61..999M} website\footnote{\url{http://maxi.riken.jp/top/index.html}}. To estimate the X-ray flux at the time of the radio observations, we downloaded the X-ray spectral and rmf files from the \maxi\ website. The data were then analysed using \texttt{xspec} (version 12.8; \citealt{1996ASPC..101...17A}) from the \texttt{heasoft} software package (version 6.21). \citet{2000ApJ...542..914W} and \citet{1996ApJ...465..487V} were used to account for line-of-sight abundances and photoionisation cross sections. The 2 -- 20 keV X-ray spectra were modelled with an absorbed power-law (tbabs $\times$ powerlaw) within \texttt{xspec}. The \nh\ was fixed to $3 \times 10^{21}$\,cm$^{-2}$ \citep{2000ApJ...542.1000B}. The photon index was left free, resulting in best fits of $\approx 2$ for all epochs \citep[in agreement with results presented by ][]{2013ApJ...767..160T}. While more complex models are required to understand the accretion geometry of NS LMXBs, the low-sensitivity of the \maxi\ data and small changes in the X-ray spectrum over this energy range meant that the data are reasonably well fit by a simple absorbed power-law ($\chi^{2} = 734$ with 685 degrees of freedom), such that they provided reasonable estimates on the 2 -- 20\,keV flux. More complex models provided similar fluxes.

\swift\ Burst Alert Telescope (\swift-BAT) Hard X-ray Transient Monitor \citep{2013ApJS..209...14K} daily count rates (15--50\,keV) were obtained from the \swift-BAT website\footnote{\url{https://swift.gsfc.nasa.gov/results/transients/}}.

X-ray count rates from 2006 could not be obtained from \textit{MAXI} as it was not yet employed. For these times, the \textit{Rossi X-ray Timing Explorer} All-Sky Monitor (\rxte-ASM) 2--12 keV light curve was extracted from the ASM Light Curves webpage\footnote{\href{http://xte.mit.edu/ASM\_lc.html}{http://xte.mit.edu/ASM\_lc.html}}.

\section{Results and discussion}

\source\ cycles between its high and low X-ray modes every $\sim$170 days \citep[e.g.,][]{1984ApJ...284L..17P,2003A&A...405..199S}. The changes in the X-ray brightness are a consequence of an evolving mass accretion rate and accretion flow structure \citep{1999ApJ...526L..33W,2003A&A...405.1033C,2007MNRAS.377.1017Z}. This source is also a known radio emitter \citep[e.g.,][]{2000ApJ...536..865F,2004MNRAS.351..186M,2017A&A...600A...8D}. In this work, we report on radio observations of \source\ taken during both its high and low X-ray modes showing a dramatic change in the observed jet emission.

\subsection{X-ray modes and origin of the radio emission}

To determine the X-ray mode of \source\ at the time of the radio observations, we inspected \maxi, \swift-BAT and \rxte-ASM light curves, which clearly show the X-ray modulations (for example, see Figure~\ref{fig:x-ray_lightcurves}). In almost all cases, \maxi\ data were used to determine the X-ray mode, where the mid-line between the maxima and minima of the periodic light curve occurs at a 2--20 keV \maxi\ count rate of $\approx$1\,\phscm. For the three 2006 ATCA observations, the X-ray mode was estimated from count rates reported by \citet{2013ApJ...767..160T}.

Due to contamination from the nearby pulsar PSR~4U1820$-$30A ($\approx$0.22\arcsecond\ away; \citealt{2017MNRAS.468.2114P}), we first explored the broadband 1--25\,GHz radio emission from the source position with multi-frequency VLA observations. Like \citet{2004MNRAS.351..186M},  we find that the emission below 4\,GHz is dominated by the nearby steep spectrum ($\alpha \sim -2$) radio pulsar, while higher radio frequencies show the flatter spectrum radio emission from \source\ (Figure~\ref{fig:radio_emission}). Hereafter, we only discuss higher frequency ($\gtrsim 4$\,GHz) radio emission associated with \source.

\subsection{An evolving radio jet between the high and low modes}

Our broadband 4--26\,GHz VLA observations of \source\ taken in 2018 during a low X-ray mode show a relatively flat radio spectrum ($\alpha = -0.25 \pm 0.18$; Figure~\ref{fig:radio_xrb}), consistent with optically-thick synchrotron emission from a self-absorbed compact jet. During the 2020 high X-ray mode the radio spectrum was steeper ($\alpha = -1.0 \pm 0.4$) and we did not detect radio emission at 22\,GHz. This steep radio spectrum is consistent with optically-thin synchrotron emission indicating that the compact jet had switched off at these higher X-ray luminosities. The observed radio emission either originated from a quenched compact jet (where we are now seeing the optically-thin synchrotron emission), or from discrete jet ejecta in a transient radio jet \citep[e.g.,][]{2001ApJ...553L..27F,2004Natur.427..222F,2006MNRAS.366...79M}. To determine if the jet changes are consistent for all high and low X-ray modes, we analysed archival VLA and ATCA radio observations (Figure~\ref{fig:radio_xrb}).

\subsubsection{Low X-ray modes}
\label{sec:hardstateradio}

In addition to our 2018 VLA observation, there were four additional ATCA radio observations taken during low X-ray modes (two in 2006 and two in 2015). \source\ was detected in both 2015 observations and one 2006 observation, exhibiting a relatively flat radio spectrum with radio spectral indices of $\alpha \sim -0.15$, consistent with a compact radio jet.

The 2015 ATCA observations of \source\ were taken during a relatively hard X-ray spectral state, i.e., toward the typical island state (IS) of atoll sources (see Figure~\ref{fig:hid_radio}). These are the first reported radio detections from this source outside of the full soft state (as described by \citealt{2013ApJ...767..160T}). Interestingly, while the jet emission may have been marginally brighter (by a factor of $\sim$1.5 -- 2) at this time (see also \citealt{2018ApJ...862...16T}), the radio spectrum ($\alpha \approx -0.15$) was consistent with other low X-ray mode radio observations. Further simultaneous radio and X-ray (spectral and timing) data are needed to accurately identify the X-ray state and to compare the radio emission. 

For the two low-mode ATCA observations taken in 2006, the observations occurred prior to the ATCA bandwidth upgrade \citep{2011MNRAS.416..832W} meaning that the sensitivity was poor and \source\ was only detected in one frequency band (detected at 8.64\,GHz but not at 4.8\,GHz) on 2006-07-23, and not at all on 2006-08-08 (Figure~\ref{fig:radio_xrb}). Using the 4.8\,GHz upper-limit and 8.64\,GHz detection implied a flat spectral index on 2006-07-23, where $\alpha > -0.1$.

\subsubsection{High X-ray modes}
\label{sec:high_modes}
For the high X-ray modes, we analysed an additional six VLA observations and two ATCA observations. In all cases where good constraints could be placed on the radio spectrum, the spectrum was steep such that $\alpha \sim -1$. 

The six VLA observations were taken over 6\,days in 2018, following an X-ray maximum as the X-ray brightness decreased. For all of these six VLA observations faint radio emission was detected from \source\ (Figure~\ref{fig:radio_xrb}). Unfortunately, the faintness of these detections meant that we were not able to identify any progressive changes to the radio emission (should they exist) between each epoch as the source evolved towards a low X-ray mode. While the measured radio spectrum of each individual epoch could have been consistent with $\sim$flat, stacking these six observations together provided much tighter constraints on the spectral index, where $\alpha = -1.25 \pm 0.45$, inconsistent (at $\approx 2.9\,\sigma$) with a flat radio spectrum.

ATCA also observed \source\ during a high X-ray mode in 2006. Unfortunately, due to the low sensitivity of those radio observations we were unable to rule out a flat radio spectrum ($\alpha = -0.6 \pm 0.8$).

\begin{figure}
\centering
\includegraphics[width=0.95\columnwidth]{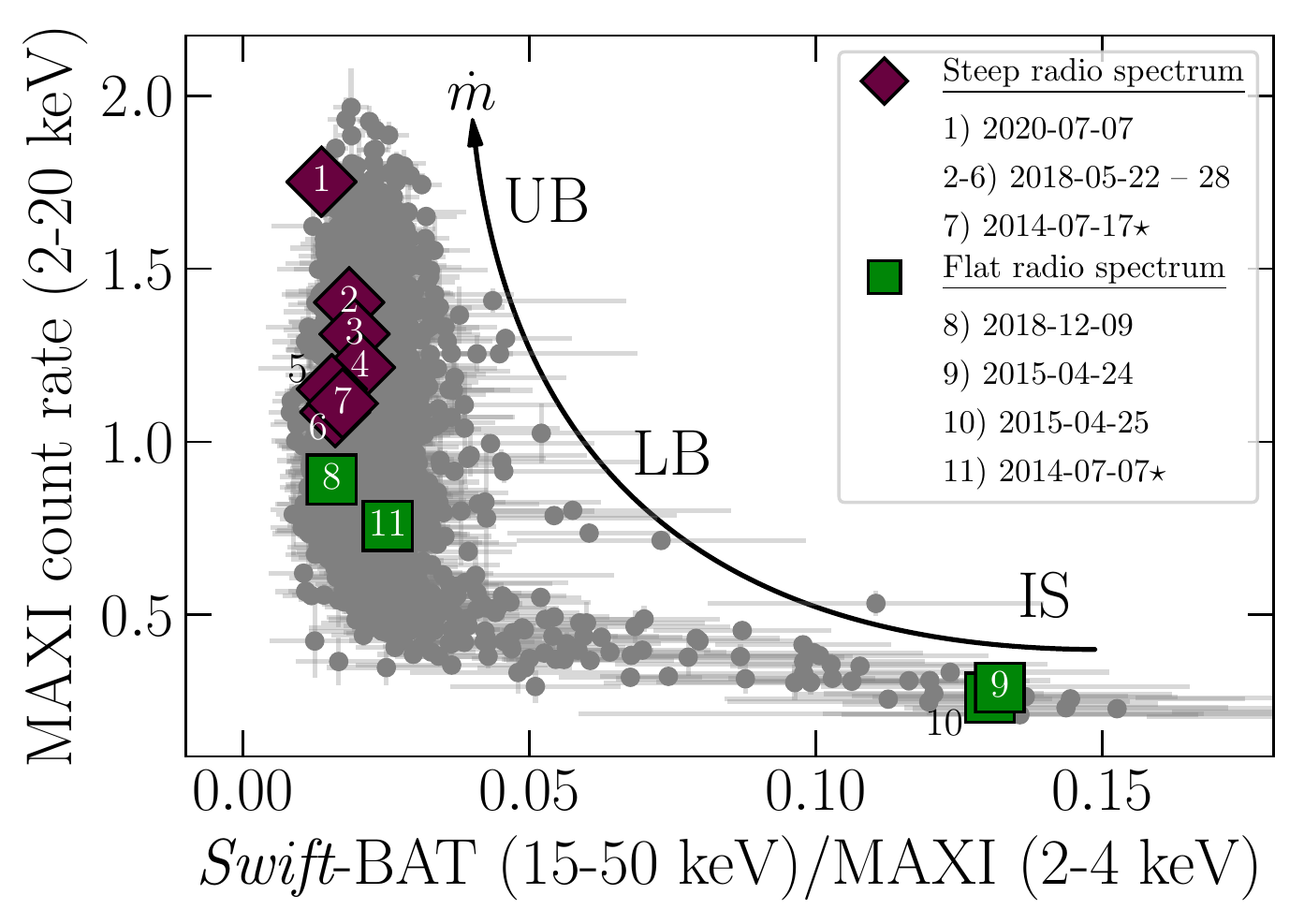}
\caption{Hardness intensity diagram of \source\ (grey markers), indicating the timing of radio observations. Magenta diamonds show radio observations with a steep radio spectrum, while the green squares represent a flat radio spectrum. The approximate location of the island (IS), lower banana (LB), and upper banana (UB) spectral branches are displayed (as depicted by \citealt{2013ApJ...767..160T}), as well as the evolution of the mass accretion rate $\dot{m}$. The numbers correspond to each observational epoch (see legend). We include the ATCA (high mode) and ALMA (low mode) 
observations from \citet{2017A&A...600A...8D}, marked with a $\star$. The 2006 ATCA detections are not shown due to \maxi\ not being available, but support the conclusions here.}
\label{fig:hid_radio}
\end{figure}

In addition, \citet{2017A&A...600A...8D} reported on ATCA (5.5/9\,GHz) and ALMA (302\,GHz) observations from 2014, where the ALMA observations were taken 10 days prior to the ATCA observations. Those observations produced puzzling results, where \source\ was detected at 5.5 and 302\,GHz, but not at 9\,GHz (despite the 5.5 and 302\,GHz results indicating that the source should have been detected at 9\,GHz). The 9\,GHz non-detection was attributed to either atmospheric or systematic effects, or even an additional previously unknown bright millimetre (mm) source close to the position of \source\ (to account for the bright 302\,GHz emission). However, analysing the \maxi\ X-ray light curves at the time of the ATCA and ALMA observations revealed that, despite occurring only 10\,days apart, the two jet observations occurred within different X-ray modes, where the ALMA observations were taken in a low mode and the ATCA taken in a high mode (corresponding to points 11 and 7, respectively, in Figure~\ref{fig:hid_radio}). Therefore, our results imply that the 9\,GHz non-detection was a result of a steep radio spectrum at the time of the ATCA observations (using the 236\,\uJybeam 5.5\,GHz radio detection and assuming an $\alpha = -0.7$ radio spectrum, \source\ should have been $\sim 167$\,\uJybeam\ at 9\,GHz, below the 200\,\uJybeam\ upper-limit reported). The standalone ALMA data from \citet{2017A&A...600A...8D} suggested a flat/inverted mm spectrum ($\alpha = 1.7 \pm 1.5$) with no strong intra-observational fading or brightening, implying emission from a compact jet. Comparing typical radio flux densities of \source\ with the bright (400\,\uJy) ALMA detection suggests a radio to mm spectral index of $\geq 0.15$ \citep[as noted by][]{2018ApJ...862...16T}.

\subsection{Quenching of the compact jet emission}

Radio observations of \source\ have shown a  dramatic change in the jet emission between the source's high and low X-ray modes. These radio observations indicate that \source\ launches a steady compact jet during its low X-ray modes but not during its high X-ray modes (Figure~\ref{fig:radio_xrb}), indicating that the jet is quenched at those times. We also constrain the timescale of the jet quenching to be less than 10\,days. For NS LMXBs, recent results have shown compact jet quenching occurring in less than 2\,days \citep{2018A&A...616A..23D}, with even shorter constraints for BH systems \citep{2020MNRAS.498.5772R}. No bright radio flaring was observed, where the observed 5\,GHz radio flux densities only varied by a factor of $\sim$2 for all observations. Therefore, while we currently favour a scenario where the high-mode steep-spectrum radio emission arises from the quenched compact jet (following the jet break evolving below the radio band as the jet's first acceleration zone moves away from the compact object; e.g., \citealt{2014MNRAS.439.1390R}), further high cadence radio observations (or sensitive high-resolution very long baseline interferometric radio observations) around the transition are required to identify whether this, or a transient jet are responsible for the observed radio emission. At present, the shortest timescale we can place on the launching, flaring, and fading of any transient jet ejecta is $\leq$10\,days (see Section~\ref{sec:high_modes}).

\subsubsection{Connection to the accretion flow}

In BH transients, compact jet quenching is typically associated with a transition from the hard to the soft X-ray spectral state. In NS LMXBs, in particular in atoll-type sources, the jet quenching scenario is not so clear, with the jet quenching observed in some sources, but not others \citep[e.g.,][for further discussion]{2011IAUS..275..233M,2017MNRAS.470.1871G}. In \source, we have observed compact jet quenching that appears to be strongly connected with the accretion rate, where the jet emission transformed at an X-ray flux of $\sim 5.1 \times 10^{-9}$\ergscm\ (3--10 \,keV), corresponding to a transition luminosity of $\sim 3.5 \times 10^{37} (D/\rm{7.6\,kpc})^{2}$\,\ergs (3--10\,keV). 

Connecting the observed source evolution with the typical atoll-type spectral branches: at its lowest mass accretion rate \source\ the X-ray spectrum is hardest (within the IS; Figure~\ref{fig:hid_radio}), then as the average mass accretion rate increases the source initially softens before rising at a constant hardness through the lower banana (LB) to the upper banana (UB) branch, with the transition of the jet occurring somewhere around the LB state (Figure~\ref{fig:hid_radio}). In atoll-type sources as the mass accretion rate rises (from the LB to UB branch), the electron temperature of the Comptonising medium (the boundary layer) remains almost constant while the optical depth increases. A change in the boundary layer, along with a simultaneous decrease of the inner disk radius \citep[e.g.,][]{2002MNRAS.337.1373G} could be responsible for the jet changes.

In comparison, in Z-type NS LMXBs, which only show soft X-ray states (but do show some changes in the X-ray hardness), the NS black-body temperature increases by a factor of $\sim$2 when $\dot{m}$ rises within the normal branch, giving an increase of the radiation pressure. Moreover, transient radio jets are launched during this progression \citep{2006MNRAS.366...79M}. In a unified model for NS LMXBs, \citet{Church2014} proposed that in both atoll and Z-type sources an increase of the radiation pressure may be responsible for disrupting the inner disk and launching transient jets, implying a connection between the jets and the inner accretion disk.

A possible coupling between the jet emission and the boundary layer is an interesting prospect. Such a connection was discussed in a study of the dwarf nova SS~Cyg, where the observed radio emission brightened around the time that material first reached the boundary layer and radio flaring was observed when the optical depth of the boundary layer increased such that it became optically-thick to itself \citep{2016MNRAS.460.3720R}. That study speculated that the boundary layer may provide a vertically-extended region required for jet launching, analogous to BH and NS LMXBs, where the WD and NS systems have a boundary layer while BH LMXBs contain a geometrically-thick accretion flow (i.e., corona; e.g., \citealt{zdziarski2004}).

\section{Conclusions}

\source\ periodically cycles between high and low X-ray accretion modes. Using new and archival radio observations of this system during each of these modes, we report on dramatic changes to the jet emission. In all observations where good constraints could be placed on the radio spectrum, \source\ launched a flat-spectrum compact jet during its low X-ray modes, which was quenched during the high X-ray modes. In the latter case, the radio spectrum was consistent with optically-thin synchrotron emission from either a quenched compact jet where the jet break had evolved below the radio band jet, or possibly a transient jet. The changes in the jet appear to be strongly linked to the X-ray luminosity, and hence the mass accretion rate, suggesting a connection to the composition of the inner accretion flow or boundary layer and not the X-ray spectral state. The observations reported here suggest that the radio emission transforms rapidly at an X-ray luminosity of $L_{\rm X (3-10\,keV)} \sim 3.5 \times 10^{37} (D/7.6\,{\rm kpc})^2$\,\ergs. High cadence radio and X-ray observations around and after the transition are required to determine precisely when the jet transition occurs, discriminate between the quenched compact jet and transient jet scenario, and identify the specific changes in the accretion flow that may be driving the jet evolution. In particular, X-ray monitoring that includes spectral and X-ray timing information is necessary.

\section*{Acknowledgements}

We thank the referee for their helpful comments. TDR, MDS and AS acknowledge financial contribution from ASI-INAF n.2017-14-H.0, an INAF main stream grant. ND is supported by a Vidi grant from the Dutch organization for scientific research (NWO). JvdE is supported by a Lee Hysan Junior Research Fellowship from St Hilda's College, Oxford. The NRAO is a facility of the NSF operated under cooperative agreement by Associated Universities, Inc. ATCA is part of the ATNF which is funded by the Australian Government for operation as a National Facility managed by CSIRO. We acknowledge the Gomeroi people as the traditional owners of the ATCA observatory site. This research made use of NASA's Astrophysics Data System, software, and HEASARC web tools, and \maxi, \swift-BAT and \rxte-ASM data.

\section*{Data Availability}

All radio results are provided in Table~\ref{tab:radio}1. VLA, ATCA, and ALMA data can be accessed from their online archives. \textit{MAXI}, \swift-BAT, and \rxte-ASM light curves are publicly available.



\bibliographystyle{mnras}
\bibliography{bib} 




\onecolumn
\begin{appendix}
\section{Online information: light curves, and radio SEDs, and radio data}

\begin{figure}
\centering
\includegraphics[width=0.6\columnwidth]{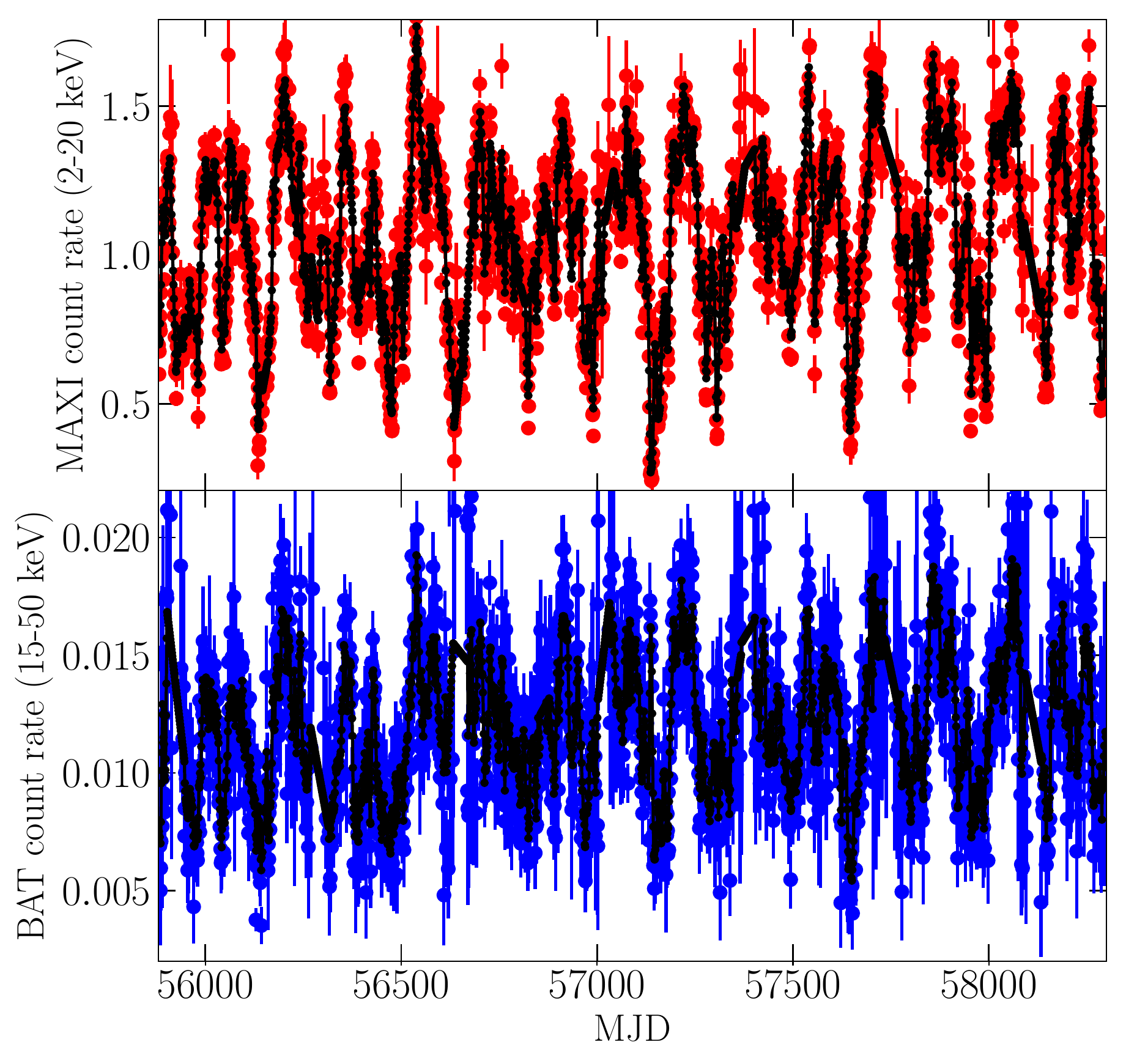}
\caption{X-ray light curves of \source, showing an example of the X-ray periodicity. The top panel shows the \maxi\ daily 2--20\,keV count rate, while the lower panel shows the \swift-BAT 15--50\,keV rate. The coloured points show the reported rates, while the black points/line gives the 5-day moving average. The X-ray modulations of \source\ are clearly visible.} \label{fig:x-ray_lightcurves}
\end{figure}

\begin{figure}
\centering
\includegraphics[width=0.6\columnwidth]{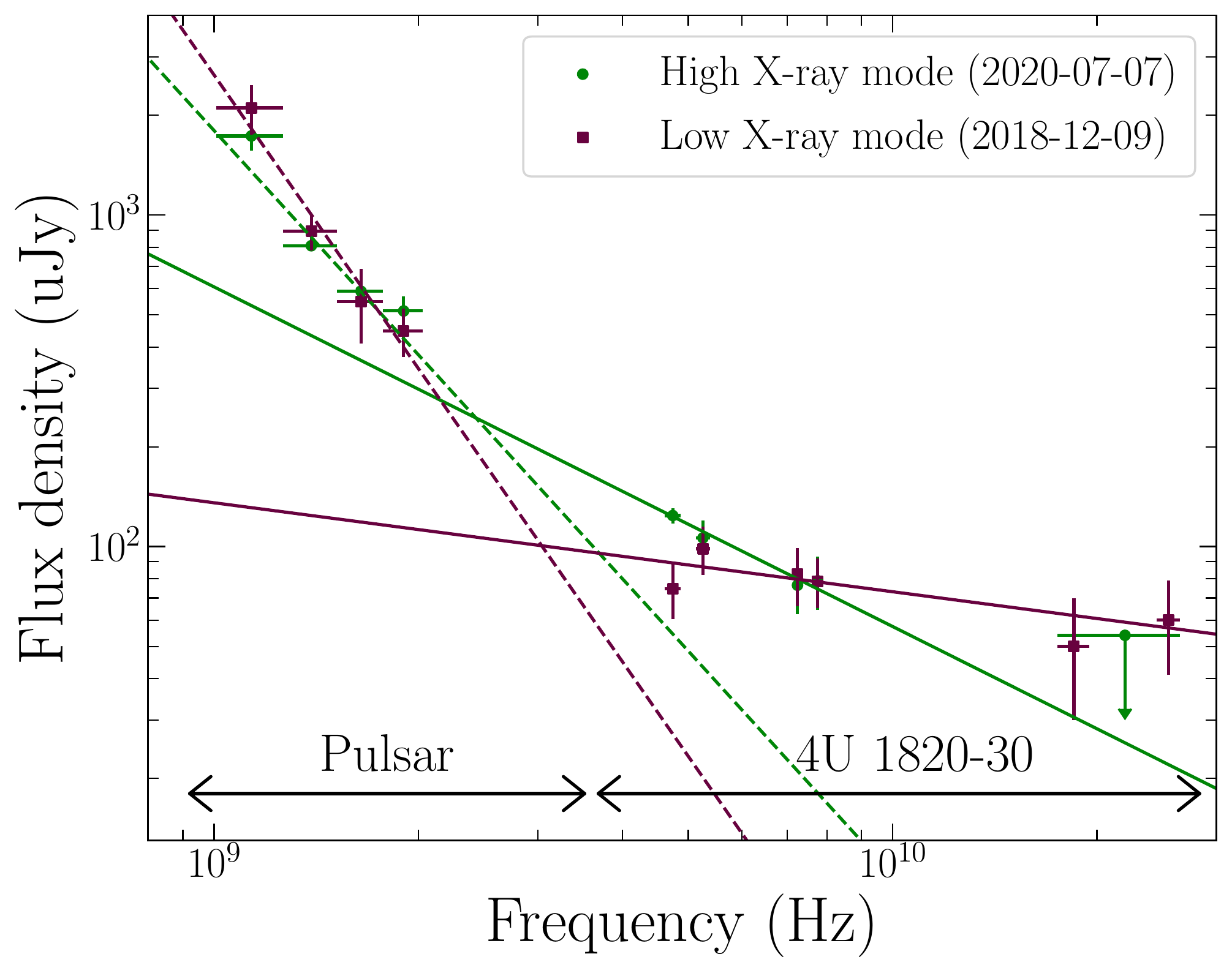}
\caption{Radio emission from the source position of \source\ showing the contribution from nearby pulsar PSR~4U1820$-$30A to the emission ($\approx$0.2\arcsecond\ away). These observations - taken during both the high (black circles) and low X-ray modes (red squares) show that \source\ dominates at frequencies above 4\,GHz, while the radio pulsar dominated the emission at lower frequencies. Fits to the pulsar data are shown by the dashed lines, while the LMXB is shown with solid lines.} 
\label{fig:radio_emission}
\end{figure}

\begin{figure*}
\centering
\includegraphics[width=0.4\columnwidth]{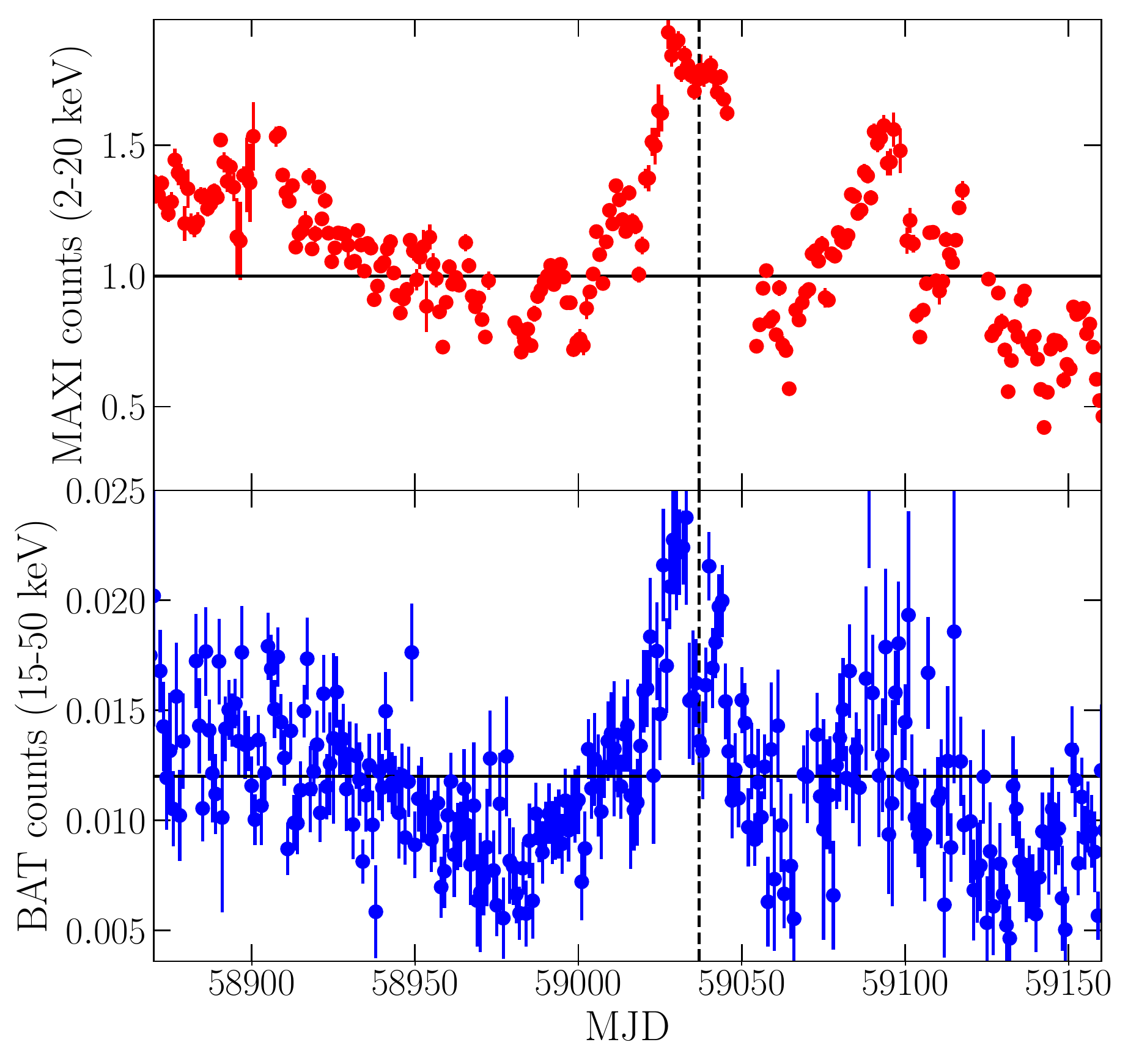}
\includegraphics[width=0.4\columnwidth]{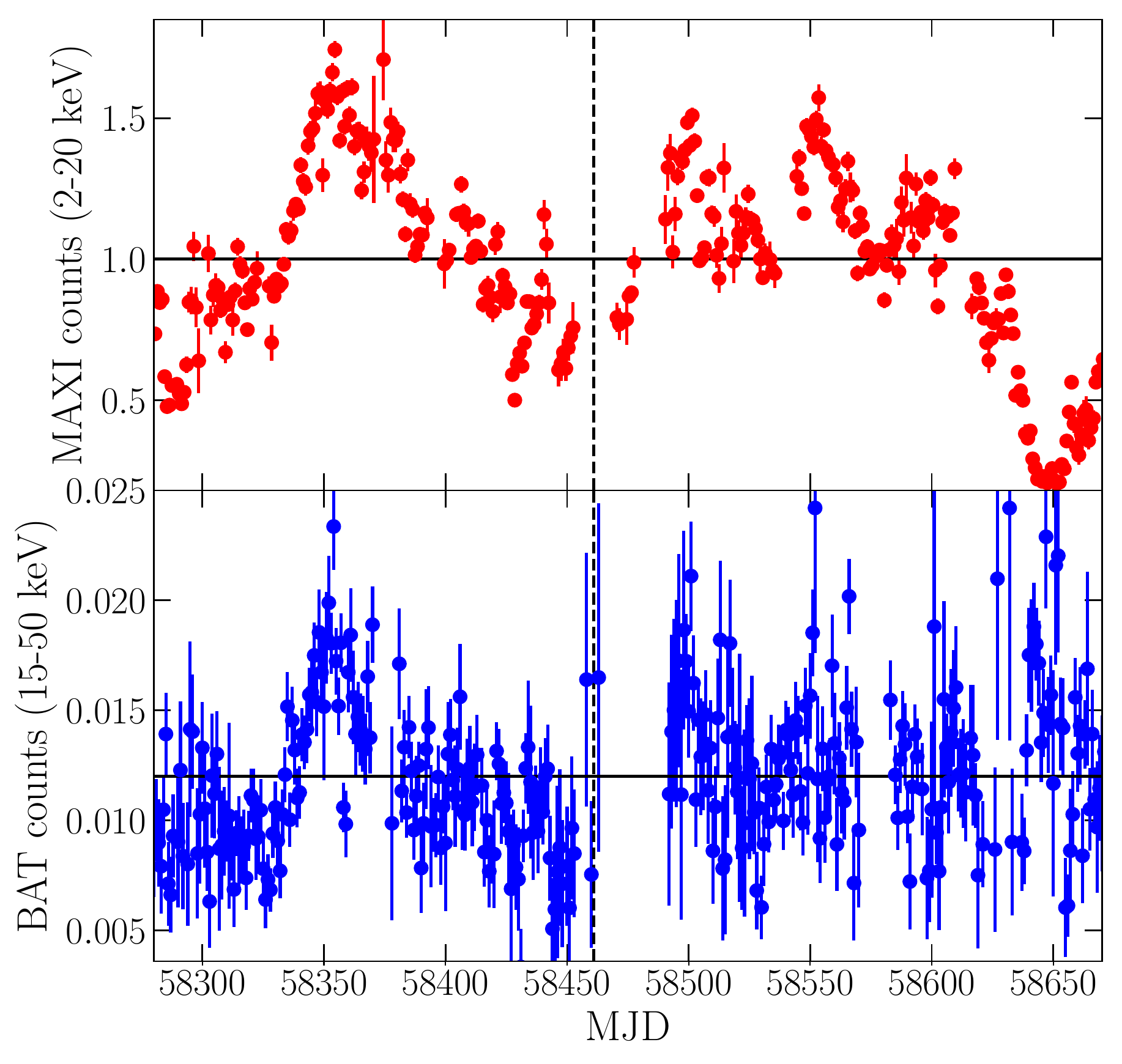}

\includegraphics[width=0.4\columnwidth]{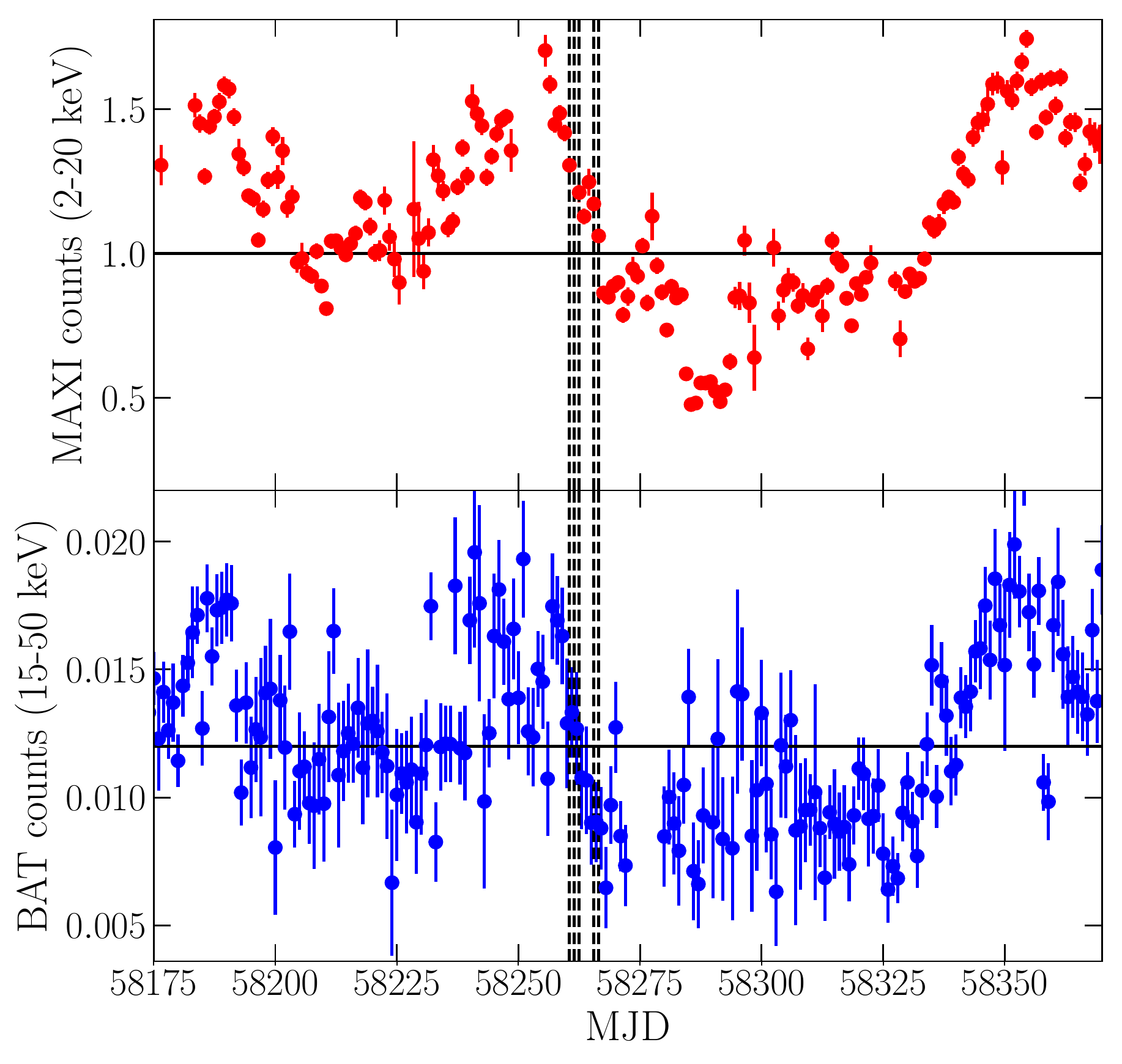}
\includegraphics[width=0.4\columnwidth]{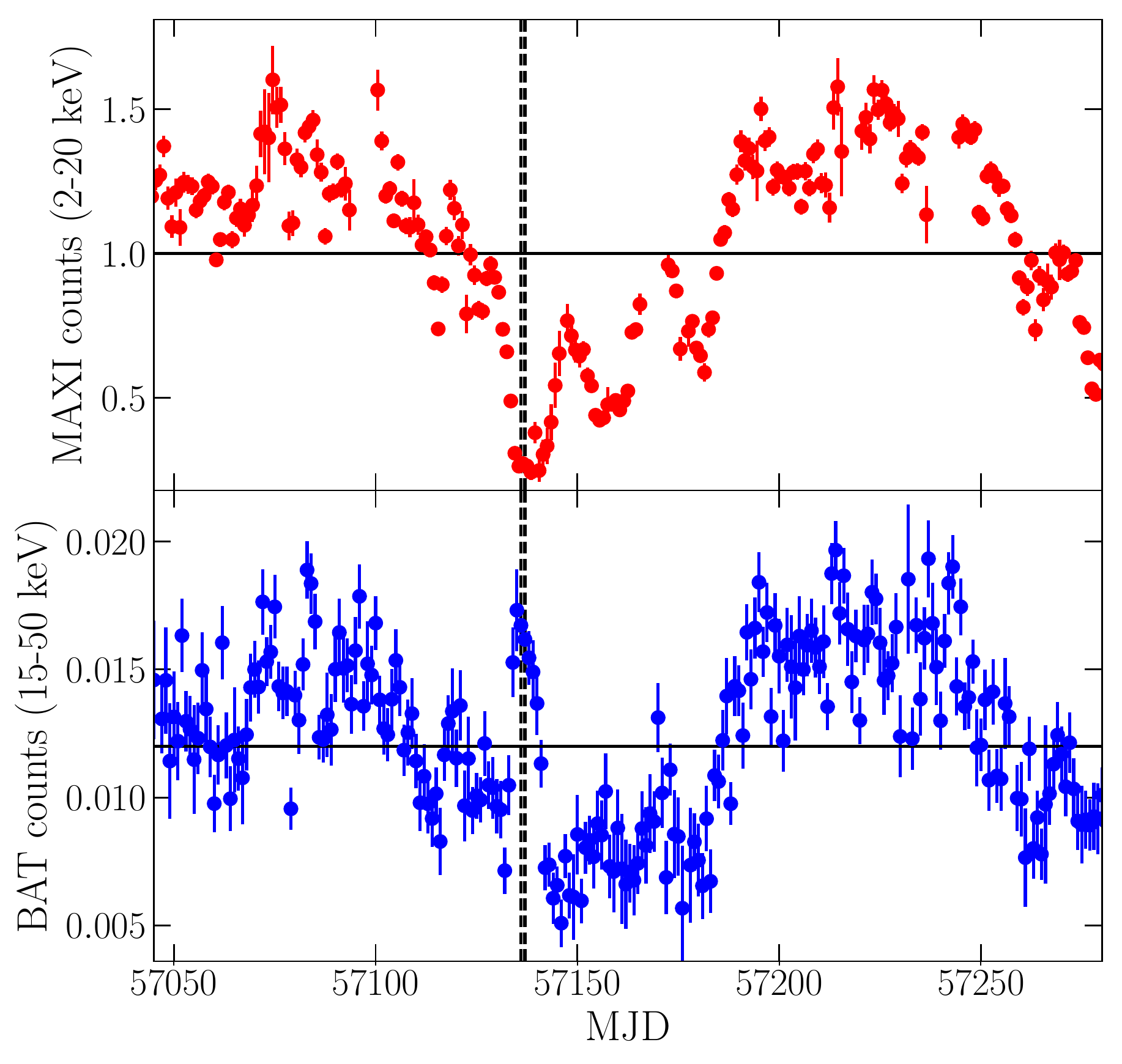}

\includegraphics[width=0.4\columnwidth]{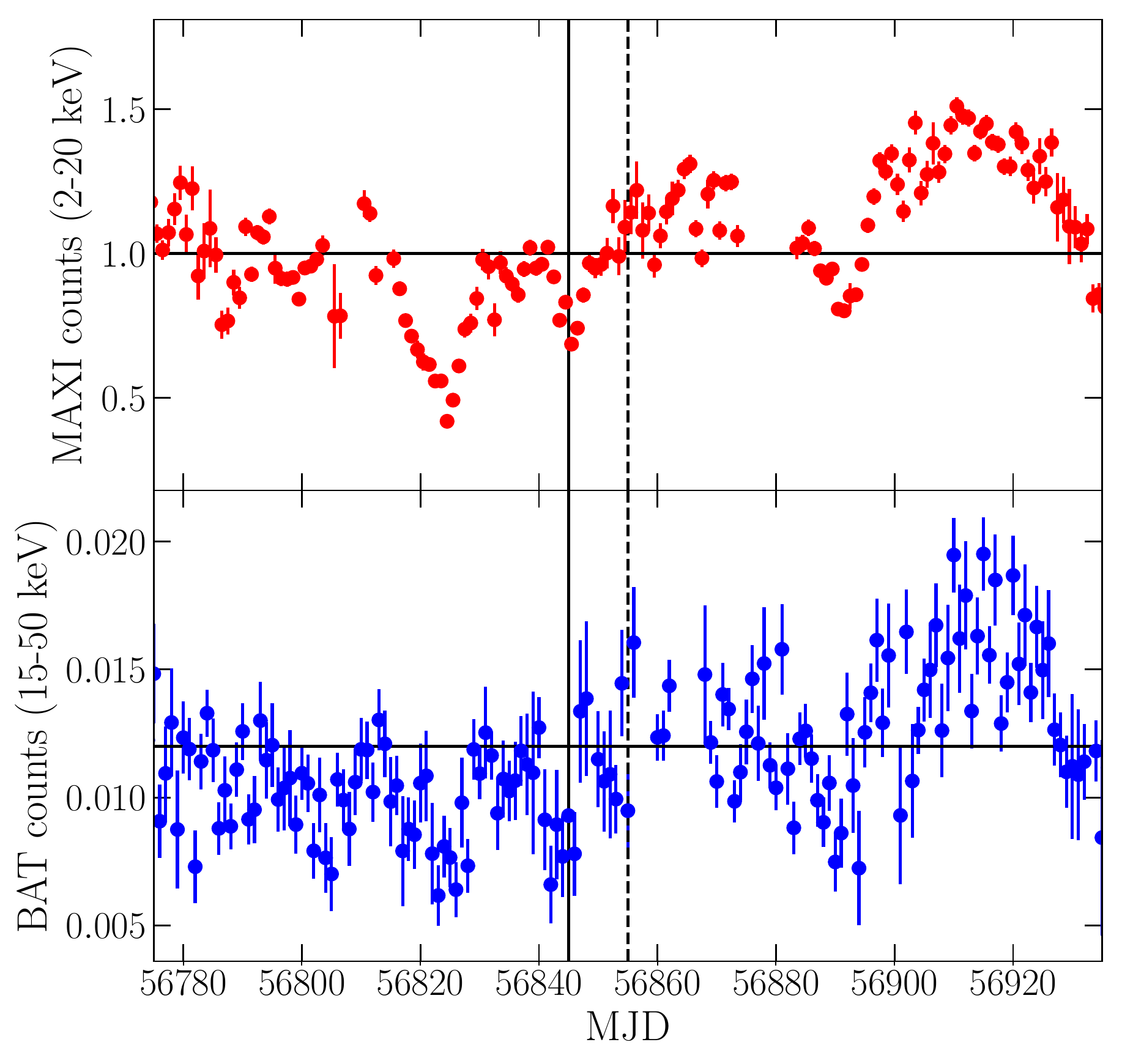}
\includegraphics[width=0.4\columnwidth]{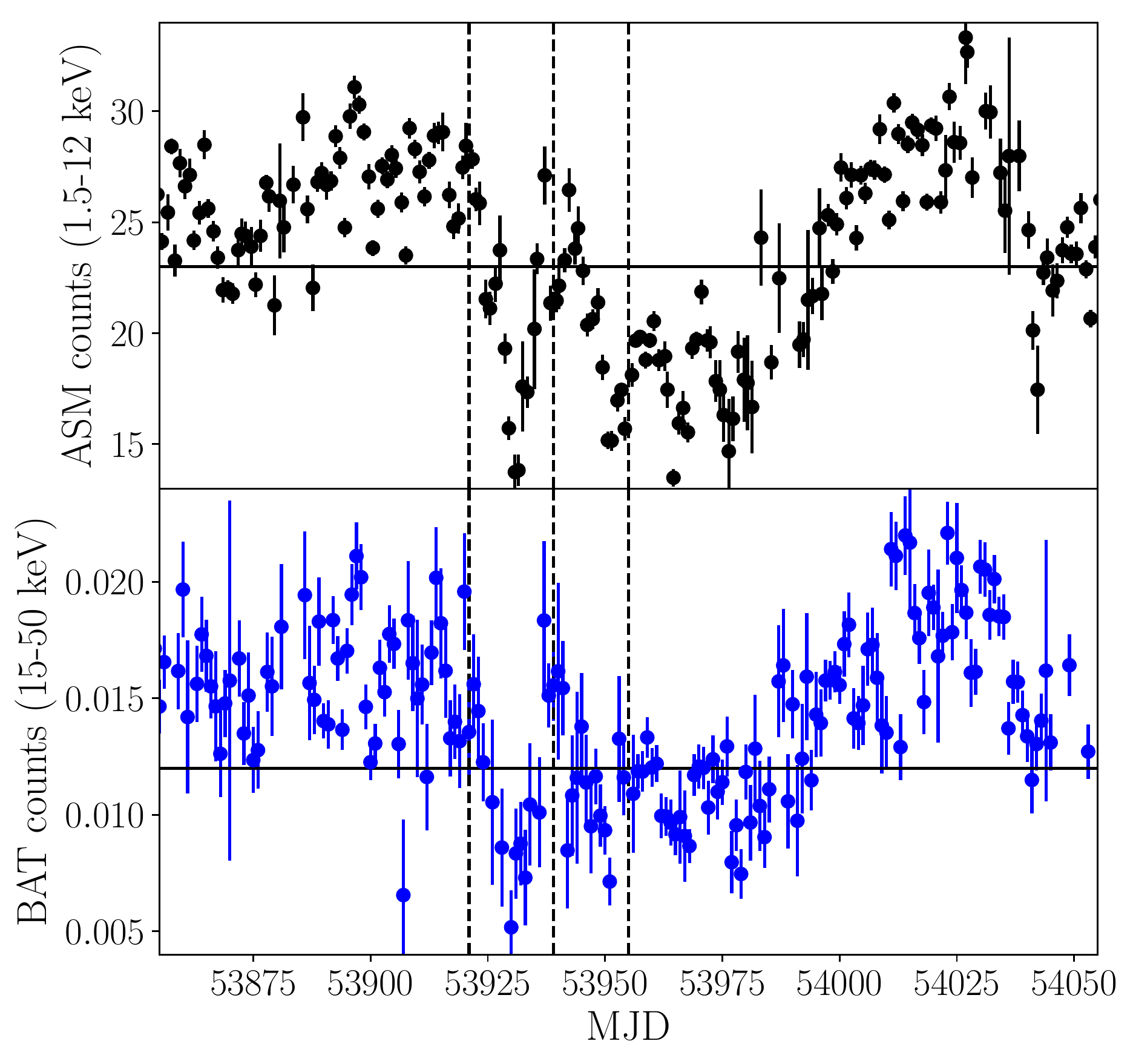}

\caption{X-ray light curves of \source\ around the timing of the radio observations discussed in this paper. The top panel of each figure shows the \maxi\ 2--20\,keV light curves, while the lower panel is the 15--50\,keV \swift-BAT light curve. Vertical dashed lines show the timing of the radio observations. Solid horizontal lines shows the approximate mid-line of the light curves. (\textit{Top left}): VLA observation taken during the 2020 high X-ray mode. (\textit{Top right}): 2018 low mode. (\textit{Middle left}): Timing of the six close-in-time 2018 high mode observations, taken as \source\ fades from an X-ray peak. (\textit{Middle right}): Radio observations taken on two consecutive days during its 2015 low mode. These two radio observations were taken in a harder X-ray state. (\textit{Lower left}): ALMA (solid vertical line) and ATCA (dashed vertical line) from 2014 taken during a low and high X-ray mode, respectively (from \citealt{2017A&A...600A...8D}). (\textit{Lower right}): \rxte-ASM and \swift-BAT light curves during the 2006 ATCA radio observations, with one high-mode (the first) and two low-mode observations (second and third).} \label{fig:lc_all}
\end{figure*}

\FloatBarrier

\begin{table*}
\label{tab:radio}
  \begin{center}
    \caption{Radio observations of \source\ during its high and low X-ray modes. Low frequency ($< 4$\,GHz) results are attributed to the nearby pulsar (PSR~1820$-$30A) and higher radio frequencies are \source. Observations taken in the high X-ray mode are presented in the top panel, while the those taken in the low X-ray mode are in the lower panel. Errors are 1-$\sigma$ and upper-limits are 3-$\sigma$. Flux density errors include systematic uncertainties.}
    \begin{tabular}{c c c c c c c c}
    \hline
          Date & MJD & Project & Frequency & Flux density & $\alpha$ & 3--10\,keV X-ray flux & X-ray mode/state \\
          & & code & (GHz) & ($\mu$Jy) &  & $\times 10^{-9}$\,erg/s & \\
      \hline
      
      2020-07-07 & $59037.22 \pm 0.03$ & 20A-255 & 1.1355 & 1730 $\pm$ 165 & $-2.2 \pm 0.3$  &  & Pulsar   \\
                 & &      & 1.3915 & 810 $\pm$ 50 &  &  &    \\
                 & &      & 1.6475 & 590 $\pm$ 70 & &  &    \\
                 & &      & 1.9035 & 515 $\pm$ 55 & &          \\

                 & &      & 4.743 & 124 $\pm$ 14 & $-1.0 \pm 0.4$  &  8.05 $\pm$ 0.20 & High/Soft \\
                 & &      & 5.255 & 106 $\pm$ 14 &  &  &  \\
                 & &      & 7.243 &  76 $\pm$ 14 & &  &   \\
                 & &      & 7.755 &  78 $\pm$ 15 & &  &      \\
                 & &      & 22.00 &  $< 54$      & &   &     \\ 
                 
2018-05-28 & $58266.457 \pm 0.015$ & 18A-081 & 8.488 & 40 $\pm$ 9 & $-2.0 \pm 1.9$ & 5.13 $\pm$ 0.15 & High/Soft  \\
                 & &      & 9.488 & 33 $\pm$ 8 &  &   &     \\
                 & &      & 10.488 & 24.0 $\pm$ 8.5 & &   &     \\
                 & &      & 11.488 & 25.0 $\pm$ 8.5 & &  &      \\

2018-05-28 & $58266.313 \pm 0.015$ & 18A-081 & 8.488 & 44 $\pm$ 8 & $-1.8 \pm 1.6$ & 5.13 $\pm$ 0.15  & High/Soft  \\
                 & &      & 9.488 & 38 $\pm$ 9 &  &   &  \\
                 & &      & 10.488 & 33 $\pm$ 9 & &   &   \\
                 & &      & 11.488 & 25 $\pm$ 8 & &   &         \\

2018-05-27 & $58265.467 \pm 0.015$ & 18A-081 & 8.488 & 38 $\pm$ 8 & $-1.8 \pm 1.5$ & 6.2 $\pm$ 0.7  & High/Soft  \\
                 & &      & 9.488 & 30 $\pm$ 8 &  &   &  \\
                 & &      & 10.488 & 31 $\pm$ 7 & &   &   \\
                 & &      & 11.488 & 26 $\pm$ 8 & &   &         \\

2018-05-24 & $58262.47 \pm 0.04$ & 18A-081 & 8.488 & 31 $\pm$ 6 &  $-0.8 \pm 1.5$  & 5.9 $\pm$ 0.2 & High/Soft  \\
                 & &      & 9.488 & 25 $\pm$ 6 &  &   &        \\
                 & &      & 10.488 & 22 $\pm$ 6 & &  &          \\
                 & &      & 11.488 & 26 $\pm$ 7 & &  &    \\

2018-05-23 & $58261.491 \pm 0.015$ & 18A-081 & 8.488 & 35 $\pm$ 8 &  $-1.4 \pm 1.3$  & 6.07 $\pm$ 0.14 & High/Soft  \\
                 & &      & 9.488 & 28 $\pm$ 6 &  &   &        \\
                 & &      & 10.488 & 22 $\pm$ 6 & &  &          \\
                 & &      & 11.488 & 24.0 $\pm$ 6.5 & &  &    \\

2018-05-22 & $58260.47 \pm 0.04$ & 18A-081 & 8.488 & 32 $\pm$ 5 & $-1.2 \pm 1.0$  & 6.05 $\pm$ 0.16 & High/Soft  \\
                 & &      & 9.488 & 28.0 $\pm$ 5.5 &  &  &          \\
                 & &      & 10.488 & 22.0 $\pm$ 5.5 & &   &  \\
                 & &      & 11.488 & 24 $\pm$ 6 & &   &         \\

2014-07-17$^a$ & $56855$ & C3010 & 5.5 & 236 $\pm$ 27 &  $< -0.1$ & 4.9 $\pm$ 0.6 & High/Soft \\
                 & &         & 9.0 & $<200$ &  &   &        \\

    2006-07-05 & $53921.57 \pm 0.24$ & C1559 & 4.8 & 170 $\pm$ 45 & $-0.6 \pm 0.8$  & 7.0 $\pm$ 0.7$^{\rm b}$ & High/Soft \\
     & &      & 8.64 & 130 $\pm$ 40 &&  & \\

\hline

      2018-12-09 & $58461.81 \pm 0.03$ & 18A-194 & 1.1355 & 2000 $\pm$ 450 & $-2.7 \pm 0.7$ &   & Pulsar \\
                 & &      & 1.3915 & 890 $\pm$ 110 &&   &      \\
                 & &      & 1.6475 & 550 $\pm$ 140 &&    &       \\
                 & &      & 1.9035 & 450 $\pm$ 75 & &     &    \\
 
                 &  &     & 4.743 & 75 $\pm$ 14 & $-0.25 \pm 0.18$& 3.76 $\pm$ 0.35  & Low/Soft  \\
                 &  &     & 5.255 & 98 $\pm$ 17 & &  &        \\
                 &  &     & 7.243 & 83 $\pm$ 17 &  &   & \\
                 &  &     & 7.755 & 79 $\pm$ 15 & &       & \\
                 &  &     & 18.5 & 50 $\pm$ 20 & &   &   \\
                 &  &     & 25.5 & 62 $\pm$ 19 & &   &   \\


    2015-04-24 & $57136.76 \pm 0.20$ & C2877 & 5.0 & 200 $\pm$ 10 & $-0.14 \pm 0.11$  & 1.23 $\pm$ 0.15 & Low/Harder \\
     & & & 6.0 & 198 $\pm$ 10 &  &  &  \\
     & & & 8.5 & 186 $\pm$ 12 &  &  &  \\
     & & & 9.5 & 184 $\pm$ 11 &  &  &  \\

    2015-04-25 & $57137.78 \pm 0.22$ & C2877 & 5.0 & 218 $\pm$ 11 & $-0.16 \pm 0.14$ & 1.20 $\pm$ 0.15 & Low/Harder \\
     & & & 6.0 & 208 $\pm$ 11 &  &  &  \\
     & & & 8.5 & 205 $\pm$ 12 &  &  &  \\
     & & & 9.5 & 190 $\pm$ 16 &  &  &  \\

2014-07-07$^a$ & $56845$ & ALMA & 295 & 390 $\pm$ 25 &  $1.7 \pm 1.5$ & 3.0 $\pm$ 0.6 & Low/Soft \\
 & & & 297 & 404 $\pm$ 26 &  &  &  \\

 & & & 307 & 415 $\pm$ 28 &  &  &  \\

 & & & 309 & 440 $\pm$ 28 &  &  &  \\

    2006-07-23 & $53939.50 \pm 0.24$ & C1559 & 4.8 & $< 126$ & $> -0.1$  & 5.26 $\pm$ 0.5$^{\rm b}$ & Low/Soft \\
         & &      & 8.64 & 166 $\pm$ 50 &  && \\
         
    2006-08-08 & $53955.50 \pm 0.23$ & C1559 & 4.8 & $< 120$ & -- & 4.7 $\pm$ 0.5$^{\rm b}$ & Low/Soft \\
                             & &      & 8.64 & $< 150$ &  && \\ 
                             
       \hline
    \end{tabular}
  \end{center}
$^{\rm a}$ From \citet{2017A&A...600A...8D}. $^{\rm b}$ From \citet{2013ApJ...767..160T}.\\
\end{table*}

\end{appendix}


\label{lastpage}
\end{document}